\documentclass[preprint,authoryear,12pt,twocolumn]{elsarticle}

\usepackage{epsfig}
\usepackage{amssymb}
\usepackage{aas_macros}

\usepackage[ps2pdf,%
a4paper=true,%
breaklinks=true,%
colorlinks=true,%
pdfauthor={First Author et al.},%
pdftitle={Template for manuscripts in Advances in Space Research}%
]{hyperref}

\journal{Advances in Space Research}
\biboptions{authoryear}

\newcommand{\mdot}{\mbox{$\dot{M}$}}
\newcommand{\Rstar}{\mbox{$R_\ast$}}

\newcommand{\vinf}{\mbox{$v_\infty$}}
\newcommand{\xmm}{{\em XMM-Newton}}
\newcommand{\cxo}{{\em Chandra}}

\newcommand{\bcep}{\mbox{$\beta$\,Cep}}

\newcommand{\zpup}{\mbox{$\zeta$\,Pup}}
\newcommand{\zori}{\mbox{$\zeta$\,Ori}}

\newcommand{\Lbol}{\mbox{$L_{\rm bol}$}}

\newcommand{\Lx}{\mbox{$L_{\rm X}$}}

\newcommand{\myr}{\mbox{$M_\odot\,{\rm yr}^{-1}$}}
\newcommand{\lsim}{\raisebox{-.4ex}{$\stackrel{<}{\scriptstyle \sim}$}}

\def\changed{}

\begin{document}

%%%%%%%%%%%%%%%%%%%%%%%%%%%%%%%%%%%%%%%%%%%%%%%%%%%%%%%%%%%%%%%%%%%%%%%%%%%%%
%% Frontmatter
\begin{frontmatter}


% Use the tnoteref command within \title and fnref within \author or \address for footnotes;
% use the corref command within \author for corresponding author footnotes;
% \author{Name\corref{cor1}\fnref{label2}}
% \address{Address\fnref{label3}}

\title{X-ray diagnostics of massive star winds}
%\tnotetext[footnote1]{This template can be used for all publications in 
%Advances in Space Research.}

% Use optional labels to link authors explicitly to addresses:

\author{Lidia M. Oskinova}
\address{Institute for Physics and Astronomy, University of Potsdam,
14476 Potsdam, Germany}
%\cortext[cor]{Corresponding author}
%\fntext[footnote2]{Additional information regarding the corresponding author}
%\ead{corresponding-author@email.address}

% Url can be given like this:
% \ead[url]{http://www.elsevier.com/wps/find/authorsview.authors/latex}

%\author{Second Author and Third Author\fnref{footnote3}}
%\address{Address of the second and third authors}
%\fntext[footnote3]{Additional information about the second and third authors}
%\ead{more@email.addresses}
%
%\author{More Authors\fnref{footnote4}}
%\address{Address of the co-authors}
%\fntext[footnote4]{Additional information about the co-authors}
\ead{lida@astro.physik.uni-potsdam.de}

\end{frontmatter}


{\noindent {\em Abstract}: Nearly all types of massive stars with radiatively 
driven stellar winds 
are X-ray sources that can be observed by the presently operating 
powerful X-ray telescopes.  In this review I briefly address recent 
advances in our understanding of  stellar winds obtained from X-ray
observations. X-rays may strongly influence the dynamics of weak 
winds of main sequence B-type stars.  X-ray pulsations were detected in a
$\beta$\,Cep type variable giving evidence of tight photosphere-wind 
connections. The winds of OB dwarfs with subtypes later than O9V may be 
predominantly in a hot phase, and X-ray observations offer the best window 
for their studies.  The X-ray properties of OB supergiants are 
largely determined by the effects of radiative transfer in their clumped  
stellar winds. The recently suggested method to directly measure mass-loss 
rates of O stars by fitting  the shapes of X-ray emission lines is considered 
but its  validity  cannot be confirmed. To obtain robust
quantitative information on stellar wind  parameters from X-ray
spectroscopy, a multiwavelength ana\-lysis by means of  stellar atmosphere
models is required. Independent  groups are now performing such analyses
with encouraging results. Joint  analyses of optical, UV, and X-ray
spectra of OB supergiants yield consistent mass-loss rates. Depending on the
adopted clumping parameters, the empirically derived mass-loss rates are  
comparable (within a factor  of a few) to those predicted by standard 
recipes \citep{Vink2001}. All sufficiently studied O stars display variable 
X-ray emission that might be related to  corotating interaction 
regions in their winds. In the latest stages  of stellar evolution, 
single red supergiants (RSG) and luminous blue variable (LBV) stars do 
not emit observable  amounts of X-rays. On the other hand, nearly all types 
of Wolf-Rayet (WR) stars are X-ray sources. X-ray spectroscopy allows a 
sensitive probe of WR wind abundances and opacities.}

\parindent=0.5 cm

%%%%%%%%%%%%%%%%%%%%%%%%%%%%%%%%%%%%%%%%%%%%%%%%%%%%%%%%%%%%%%%%%%%%%%%%%%%%%
%% Main text
\section{Introduction}

Stars much heavier than the Sun ($M_{\rm initial}>8\,M_\odot$) are
extremely  luminous and drive strong stellar winds, blowing a large part
of their matter  into the galactic environment before they finally
explode as a supernova. By  this strong feedback, massive stars ionize,
enrich, and heat the interstellar  medium, regulate the star formation,
and affect the further development of  the cluster in which they were
born. Empirical diagnostics of massive star spectra  provide quantitative 
information about stellar and wind parameters, such as the terminal
wind velocity, $v_\infty$, and the mass-loss rate, $\dot{M}$. 

With the development of space based telescopes, the classical analyses 
of optical, and radio radiation was extended to the ultraviolet 
(UV) and the X-ray range. The optical range is the easiest to access. However 
in the vast majority of OB stars the optical spectra are dominated by 
photospheric lines.  Only the most luminous stars 
with strongest stellar winds (such as Wolf-Rayet (WR) stars) show many wind 
lines in emission in the optical.  OB stars commonly 
display wind signatures in their UV  spectra, but UV  observations 
are scarce. For main-sequence B stars the wind 
signatures are marginal and difficult to disentangle from the  photospheric  
spectra even in the UV. In X-rays, on the other hand, we can observe 
emission lines from winds of  nearly all types of massive stars, including  
B-type dwarfs and other stars with weak winds. 
Thus, X-rays provide an excellent, sometimes unique  wind diagnostic.  

The history of massive star X-ray astronomy begins with the UV 
observations. Back in the 1970's the observatory {\em Copernicus} 
made the important  discovery of strong lines of highly ionized  ions  such as 
O\,{\sc vi}, N\,{\sc v},  and C\,{\sc iv} in the UV spectra of massive stars as 
cool as spectral type B1. It became immediately clear 
that the stellar effective temperatures are not sufficiently high to power 
such high degrees of ionization. E.g.\ \citet{wrh1981} showed that the lines of 
O\,{\sc vi} and N\,{\sc v} observed in the UV  spectra of the B-type star 
$\tau$\,Sco  cannot be reproduced by the standard wind 
models.  While different theories were put forward to explain the presence of 
high ions in stellar spectra, it was the work of \citet{co1979} 
where the Auger ionization by X-rays was suggested as an explanation. 

The first X-ray  observations by the {\em Einstein}  observatory indeed 
detected X-rays from OB stars, and hence proved the importance of the Auger 
process for stellar winds \citep{Seward1979,Harnden1979}. One of the 
instruments on board of the {\em Einstein} observatory was the Solid State 
Spectrometer (SSS) that was used to observe O-type stars. 

Already at these early days of X-ray astronomy, \citet{StewartFabian1981} 
employed {\em Einstein} spectra to study  the transfer of X-rays 
through a uniform stellar wind as a mean to determine stellar mass-loss rates. 
They applied a photoionization code to calculate the wind  opacity. Using 
\mdot\ as a model parameter, they found from matching the model  and the 
observed X-ray spectrum of \zpup\ (O4I) that the X-ray based mass-loss rate is 
lower by a factor of a few than obtained  from  fitting the H$\alpha$ 
emission line
and the radio and IR excess. As the most plausible explanation for this
discrepancy they suggested that the mass-loss rate derived from 
H$\alpha$ is overestimated because of wind clumping. In retrospect, 
this was a deep insight confirmed by follow-up studies only in the 21st 
century. 

These first low-resolution X-ray spectra of OB stars also showed emission 
signatures  of such high ions as S\,{\sc xv} and Si\,{\sc xiv}, and thus 
revealed the  presence of plasma with temperatures in excess of a few MK.   
From this moment on, a quest to explain X-rays from stellar winds has began. 
Among the first proposed explanations was a picture where the stellar wind  
has a very hot base zone where the plasma is constrained by a magnetic field. 
X-rays are produced in this base corona, and ionize the overlaying 
cool stellar wind. Models predict that  X-rays originating 
from this inner part of the winds should become strongly absorbed in the 
overlaying outer wind. Therefore, when observations showed only 
little absorption of X-rays, the base corona model was seriously 
questioned \citep{Long1980,Cas1981,cas1983}. A 
way to resolve this ``too little absorption'' problem was shown 
only much later (see discussion in Section\,\ref{sec:xprof}).

Besides the magnetically confined hot corona scenario, other models were put 
forward to explain X-ray emission from  single massive stars.  Detailed 
(albeit 1-D) radiative hydrodynamic models predict the development of strong 
shocks \citep{Owocki1988, felda1997, Run2002} as a result of the line driving 
instability (LDI) intrinsic to radiatively driven stellar winds 
\citep{LucyW1980, Lucy1982}. A fraction of otherwise cool ($T_{\rm cool}\sim 
10$\,kK) wind material is heated in these shocks to a few MK, and cools 
radiatively via X-ray emission. In the hydrodynamic model of \citet{feld1997},  
the X-rays are generated when a fast parcel of gas rams into a slower-moving 
dense shell, both structures resulting from the LDI. This model was successful 
in quantitatively explaining the observed low-resolution {\em Rosat} spectrum 
of an O supergiant, and is often invoked as the standard scenario for the 
origin of X-rays in stellar winds.    

Another family of models suggests that radiatively driven blobs of 
matter plough through an ambient gas that is less radiatively accelerated  
\citep{Lucy2012,Guo2010}. These blobs might be the result of an  
instability \citep{LucyW1980},  or seeded at or below the stellar  
photosphere \citep{Waldron2009,Cant2011}. When these blobs propagate,  
forward shocks are formed, where gas is heated giving
rise to  X-ray emission. \citet{Cas2008} \& \citet{Ignace2012} calculated 
the temperature and density in such a bow shocks to 
interpret some major X-ray properties: the power-law distribution of the 
observed emission measure derived from many hot star X-ray spectra, and the 
wide range of ionization stages that appear to be present throughout 
the winds. One  can also envisage a ``hybrid'' scenario to explain 
the  X-ray emission from massive stars by a combination of magnetic mechanisms  
on the surface with shocks in the stellar 
wind \citep[e.g.][]{cas1983,Waldron2009}. 

\begin{figure}[t]
\centering
\includegraphics[width=\columnwidth]{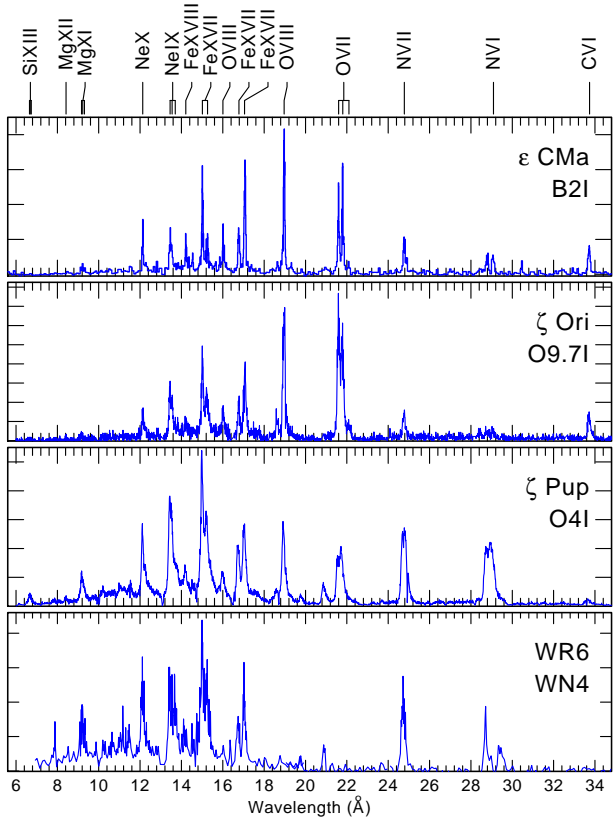}
\caption{A sample of high-resolution X-ray spectra obtained with the RGS 
spectrograph on board of \xmm. The vertical axis is the X-ray count rate in 
arbitrary units. Strong lines are identified at the top.    
  \label{fig:zxmm}}
\end{figure}

These and alternative models for the X-ray production in stellar winds are now 
rigorously checked by modern observations. The 21st century's  X-ray telescopes 
\cxo\ and \xmm\ made high-resolution 
X-ray spectroscopy possible \citep{Brin2000,rgs2001,Can2005}. 
On  board of \cxo\ is the High Energy Transmission Grating 
Spectrometer (HETGS/MEG) with a spectral resolution of $0.024$\AA\ in its 1st 
order. The Reflection Grating  Spectrometers (RGSs) of \xmm\ have a more modest 
spectral resolution of $0.05$\,\AA, but higher sensitivity. 
Figure\,\ref{fig:zxmm} shows examples of high-resolution X-ray spectra. Both 
telescopes also provide a possibility for low-resolution X-ray spectroscopy, 
imaging, and timing analysis.

%============================ Figure  RGS ===============================
%=======================================================================

In this  review I briefly address the diagnostic potential of X-rays for our
understanding of winds from single stars. General X-ray properties of 
massive stars are introduced in Section\,\ref{sec:lxlbol}, and the location of 
X-ray plasma in their winds is discussed in Section\,\ref{sec:fir}.
The classical UV diagnostics of stellar winds and the influence of X-rays 
on these diagnostics are considered in Section\,\ref{sec:drive}. The role of 
X-rays in resolving the so-called ``weak wind problem'' is discussed  in 
Section\,\ref{sec:bw}. The X-ray properties of B-stars, including pulsating 
$\beta$ Cep-type variables and magnetic stars, are briefly considered in 
Section\,\ref{sec:b}. Section\,\ref{sec:xvar} deals with using  X-ray 
variability to study the structure of stellar winds. Section\,\ref{sec:obi}
introduces modeling approaches for the description of X-ray line spectra. 
The diagnostic power of X-ray emission lines is discussed in 
Section\,\ref{sec:xprof}. Modern approaches combining X-ray spectroscopy with 
a multiwavelength analysis are presented in Section \ref{sec:nlte}. X-ray 
diagnostics of massive stars in the latest stages of their 
evolution, RSG, WR, and LBV stars, are considered in Section\,\ref{sec:wr}.
A summary concludes this review  (Section\,\ref{sec:sum}).

\section{General X-ray properties of massive stars}
\label{sec:lxlbol}

Already in the  dawn of stellar  X-ray  astronomy it was noticed that the   
X-ray luminosities of O stars   roughly correlate with their bolometric 
luminosities as  $\Lx \sim 10^{-7}  \Lbol$
\citep{Long1980,Cas1981,Pallavicini1981,Berghoefer1997}. Follow-up 
investigations largely confirmed this empirical correlation and
established that  it holds not only for single O stars but also for
binaries  \citep{osk2005,naze2009}. 

The $\Lx \sim 10^{-7}  \Lbol$ correlation holds over a limited range of 
spectral types, and apparently does not extend to stars with weaker nor 
with stronger winds than O-type stars.   
E.g.\ in their studies of the Orion  Nebula Cluster (ONC), \citet{Stelzer2005} 
showed that the X-ray properties of late O  and early B dwarfs are diverse, 
with luminosities that differ by orders of magnitude from the correlation 
$\Lx \sim 10^{-7} \Lbol$. On the basis of their studies of the star cluster 
NGC\,6231,  \citet{Sana2006} noticed that the X-ray emission from B stars 
cannot be considered as a continuation of the O star X-ray properties to lower 
luminosities. Hardly any  correlation between \Lx\ and \Lbol\ was found 
for B stars in the Carina Nebula \citep{Naze2011}. 
The X-ray luminosity  steeply drops at spectral types later than B2 
\citep{Cas1994}.  Interestingly, this decline occurs  roughly at the minimum 
bolometric luminosity required to initiate and  sustain a radiatively driven 
stellar wind \citep{Abbott1979}. 
On the other hand, stars with strong winds, such as single WR stars, also do 
not follow an $L_{\rm X}\propto L_{\rm bol}$ correlation 
\citep{Wes1996,Ignace1999,Ignace2000,osk2005}.

The physics behind $\Lx \propto  \Lbol$ is not  yet understood. 
\citet{Chleb1991} found a dependence between the wind energy of the 
single stars and their X-ray luminosity, and suggested that this 
dependence is at least partially responsible for the proportionality between 
\Lx\ and \Lbol. \citet{oc1999} and \citet{Ignace1999} suggested that a specific 
radial distribution of the X-ray filling factor may explain the $L_{\rm 
X}\propto L_{\rm bol}$ correlation for OB stars, and the
lack of such correlation for WR stars. \citet{osk-an2011} considered the
possibility that the X-ray  luminosity is correlated  with the stellar
parameters via a dependence of  the properties of the subphotospheric
convective zone  on   $L_{\rm bol}$ and $T_{\rm eff}$. E.g.,
\citet{Cant2011} predict  that surface  magnetic fields are common in
OB stars, and their  strength is increasing for hotter and more luminous
stars.  The wind momentum is also increasing with luminosity
\citep{CAK1975,puls2006}.  Thus, the ratio of the wind kinetic energy to the
magnetic energy could  remain similar for stars of different
luminosities. Interestingly,  \citet{Walborn2009} found from a study
of high-resolution   X-ray spectra of O star that the X-ray plasma
temperature correlates with the stellar effective temperature, indicating a 
close association between the hot plasma  and the stellar (sub)photosphere. 

On the other hand,  \citet{owocki2013} further investigated the idea
that the $\Lx \propto L_{\rm bol}$ correlation is secondary to a
correlations between $\Lx$ and stellar wind properties. In their study
of radiative shocks, it was found that accounting for the thin-shell 
instability \citep{Vish1994} leads to a correlation $\Lx \propto 
(\dot{M}/v_\infty)^{1-m}$, where $m$ is some exponent.  Making the 
assumption $m \approx 0.4$ results in the observed linear 
$\Lx\propto \Lbol$  correlation. This model also provides a natural explanation 
why stars with thin winds, such as B dwarfs, and stars with thick winds, such 
as WR stars, have a different scaling between X-ray and bolometric luminosity 
than the O-type stars (see review by K. Gayley in this volume). Future work 
should show how realistic is the mixing exponent ansatz. 

{\changed The X-ray spectra of single massive stars are, in general,  well 
described 
by optically thin multi-temperature plasma in collisional equilibrium. The 
spectra are comparably soft -- the emission measure weighted mean temperature 
usually does not exceed a few MK 
\citep[e.g.][]{Pallavicini1989,Zhekov2007,naze2009}. The X-ray emission is 
remarkably constant on short time scales of a few hours, but typically 
shows non-coherent slow fluctuations on a few percent level and on the time 
scale of days. While the former time scale (hours) is comparable to the 
dynamical time of the stellar wind, the latter time scale (days) is likely 
connected with stellar rotation.     

To summarize, a normal massive OB star is a nearly constant (not flaring) 
X-ray source with a luminosity $\sim 10^{-7}\,L_{\rm bol}$ and a thermal X-ray 
spectrum with an average temperature of a few MK.}

\section{Location of the hot X-ray emitting plasma in OB star winds}
\label{sec:fir}

\begin{figure}[t]
\centering
\includegraphics[width=\columnwidth]{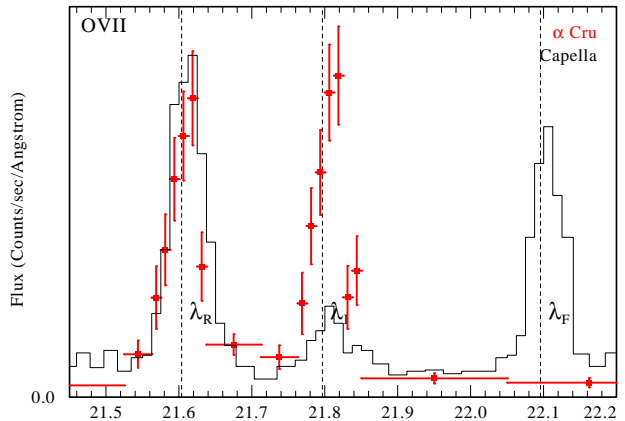}
\caption{The O\,{\sc vii} {\it fir} triplet ($\lambda_{\rm w} = 21.60$\,\AA, 
$\lambda_{\rm xy} = 21.8$\,\AA, $\lambda_{\rm z} = 22.098$\,\AA) in the 
\cxo\ LETGS spectra of $\alpha$\,Cru (B0.5IV) (red symbols) and Capella 
(G5III) (black histogram). The central wavelengths of the {\it r}, {\it i}, 
and {\it f} lines are indicated by dashed vertical lines. The {\it r} lines 
from both stars are normalized to the same height. In the spectrum of 
$\alpha$\,Cru the forbidden line is strongly suppressed due to the effect of 
strong stellar UV radiation. Based on our \cxo\ observations of $\alpha$\,Cru 
and the archival \cxo\ data of Capella.}   
\label{fig:acruox}
\end{figure}

High-resolution spectroscopy allows to constrain the location of hot emitting 
plasma. This can be done by analyzing spectra of He-like ions. which display 
prominent lines in the X-ray spectra of massive stars (see 
Fig.\,\ref{fig:zxmm}). The spectrum of a He-like ion 
consists of a resonance line,  {\it  r},  intercombination lines {\it i}, 
and a highly forbidden line {\it f}. The ratios between the fluxes in these 
lines are strongly sensitive  to the conditions in the emitting plasma.  
For instance, the ratio of fluxes between the forbidden and  intercombination 
components, ${\cal R}$, is sensitive to the electron  density and the 
ultraviolet flux \citep{Blum1972, porq2001}. The UV radiation suppresses  
the forbidden line and enhances of the intercombination line, mimicking
the effect of higher electron density. In typical OB-type stars, their copious  
UV radiation is the dominant cause for forbidden line depopulation 
(Fig.\,\ref{fig:acruox}).
 
%
%---------------------------------------------------------------------------
%-----------------------------------------------------------------------------
%

{\em If} the stellar UV photospheric flux, $H_{\nu}$, is known, the geometrical 
dilution factor $W$ can be inferred from measured value ${\cal R}(WH_\nu)$. 
This constrains the radius $R_{\rm fir}$ where the X-ray emitting plasma is 
located \citep[e.g.][]{wc2001,Leu2006,Shenar2015}.  

One of the first surprises brought by the {\em Chandra} and \xmm\ 
spectroscopy of OB stars was the discovery that the hot plasma is  
located not far out in the wind, but instead resides already  
very close to the stellar photosphere  \citep{wc2001,Kahn2001}. 

Furthermore, \citet{WC2007} pointed out a ``high-ion near star problem'': 
lines of ions with higher ionization potential are formed closer to the 
stellar surface than those of lower ions.
Particularly, the line formation regions of Si\,{\sc xiii}
and S\,{\sc xv} in the O-supergiant \zori\ (O9.7Ib) are  very close to the 
stellar photosphere, where X-rays emission is not expected  theoretically 
\citep{feld1997,Krta2009}. The proposed explanation invoked surface magnetic 
fields. Soon after, a surface magnetic field  was indeed detected on \zori\ 
\citep{Bouret2008}. Interestingly, \citet{berg1994} reported an  X-ray flare 
from this star. The recent development in stellar structure modeling shows 
that small-scale surface magnetic fields may be common in  massive
stars \citep[e.g.][]{Cant2011}.

From the analysis of a sizable sample of OB star X-ray spectra, 
\citet{WC2007} found a good correlation between the {\em f/i}-inferred 
radii and the radii where the X-ray continuum optical depth is unity.
This provides evidence for the absorption of X-rays in the stellar wind, 
in agreement with theoretical expectation.   

Recently, large progress was made in improving the {\em f/i} line ratio 
diagnostics by providing a more realistic description of the UV radiation field 
in the stellar wind. For wavelengths in the observable part of the UV 
spectrum, the flux $H_{\nu}$ can be, in principle, directly inferred from 
observations. Unfortunately, especially for giant and supergiant stars, the 
photospheric fluxes are
often contaminated by wind lines. E.g.\ the O\,{\sc vi} resonance
doublet overlaps with the $\lambda_{{\rm f} \rightarrow {\rm i}}$
transition at 1033\,\AA\ for Mg\,{\sc xi}.  The O\,{\sc vi} doublet
itself can only be reproduced by stellar atmosphere models if the X-ray
field causing Auger ionization is included  \citep{osk2006}.  This
example highlights the need of a consistent radiative transfer 
treatment for the correct interpretation of the {\em fir} line ratios.
E.g.\ \citet{osk2012}, \citet{Herve2012}, \citet{Shenar2015}, 
\citet{Rauwa2015}, \citet{Puebla2016} independently used sophisticated NLTE 
stellar atmosphere codes to predict the {\em f/i} line ratio in 
dependence on radius. Empirical {\em f/i} line ratios can be  measured 
from the observed high-resolution X-ray spectra of OB stars and compared to 
the model predictions, allowing to constrain the plasma location. Comparison 
between the model and observations show that X-ray emission in OB 
supergiants has onset radius  below $\sim 1.5$\,$R_\ast$ and is distributed 
over a large range of radii. These findings strongly constrain the models of 
X-ray production in massive stars. 

\section{Influence of X-rays on stellar wind ionization balance: including 
X-rays in the UV wind diagnostics}
\label{sec:drive}

The usual empirical diagnostics of winds from OB stars is by analysis of  UV 
resonance lines of abundant metal ions. By comparing observed and model lines 
of an ion in some ionization stage, the product of mass-loss rate and the 
ionization fraction is derived. As a next step, an assumption about the 
ionization fraction is made, e.g.\ it may be assumed that the given ionization 
stage is the leading one. Based on this assumption, the mass-loss rate can be 
evaluated. 

A puzzling result was obtained when \citet{Prinja1989} applied this method, 
to  the UV high-resolution spectra of 40  non-supergiant B stars obtained with 
the {\em IUE} observatory. It was found that, in  general, wind velocities are 
in the range of a  few$\times 100$\,km\,s$^{-1}$ and do not exceed the 
photospheric escape velocity. It was also found that the wind column densities 
are not significant -- the product of mass-loss rates and ionization fractions 
of leading ions are small,
e.g.  $\log{(\dot{M}q({\rm C\,{\sc III}}))}< -10$\,[$M_\odot$\,yr$^{-1}$]. 

\begin{figure}[t]
\centering
\includegraphics[width=\columnwidth]{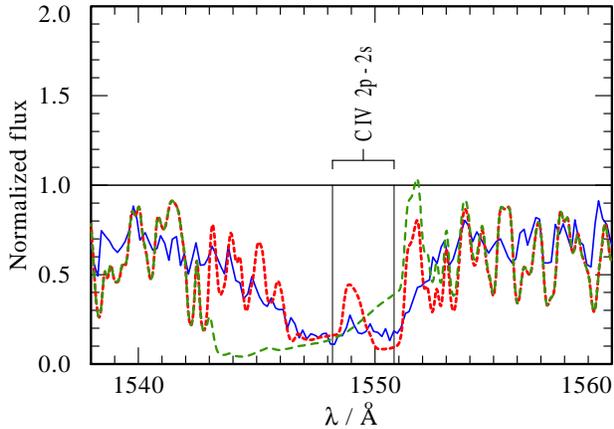}
\caption{ The effect of ionization by X-rays on the C\,{\sc iv}
 $\lambda\lambda$1548.2,\,1550.8\,\AA\ doublet.
Detail of the UV spectrum of $\tau$ Sco (B0.2V) as observed with IUE (blue
  thin line) vs.\ PoWR models: without X-rays (green line) and
  with X-rays (red line). The model parameters:
  $\log(\dot{M})=-9.3$, $v_\infty = 1000$\,km\,s$^{-1}$. This figure
  shows that C\,{\sc iv} is efficiently destroyed by X-rays in
  the outer parts of the atmosphere. Without accounting for the
  ionization by X-rays, the mass-loss rate would be underestimated. 
  Adopted from \citet{osk2011}. 
  \label{fig:tscoX}}
\end{figure}

However, the presence of X-rays changes the wind ionization balance and, 
hence, may strongly affect the UV wind diagnostics  \citep[e.g.][see also the 
review by Krticka \& Kubat in this volume]{Waldron1984,Mac1994,Massa2003}. 
\citet{Waldron2010} pointed out that if X-rays are not included in the 
models, the ionic fractions could be incorrectly estimated leading to errors 
in the derived mass-loss rate. Figure\,\ref{fig:tscoX} demonstrates the 
influence of X-rays on the empirical mass-loss diagnostics based on the 
modeling of UV  resonance lines. In $\tau$ Sco, the observed C\,{\sc iv} 
doublet can be  reproduced without X-ray superionization only by assuming a 
very low mass-loss rate. On the other hand, including X-rays at the observed 
level in the models  improves the fits and allows to reproduce the 
C\,{\sc iv} doublet with a higher mass-loss rate. This example shows 
that X-rays  should be included in the stellar atmosphere models  to 
obtain  correct  empirical mass-loss rates from the analysis of UV
spectra. 

Such analysis was done by \citet{osk2011}, who included the X-ray field 
in the detailed modeling of the UV  spectra of a sample of non-supergiant 
B stars. It was shown that the models are in agreement with empirical results 
of \citet{Prinja1989} -- the mass-loss rates in  main-sequence B stars are 
low, with $\log{\dot{M}}<9$, and the wind velocities are of the order 
of the wind escape velocity. 

%============================ Figure  ===============================
%=======================================================================

The influence of X-rays on the ionization structure of stellar winds 
was also considered as a possible solution for the very low mass-loss 
rates derived from modeling UV resonance lines, 
such as e.g. P\,{\sc v} $\lambda\lambda 1118,1128$\,\AA\ resonance 
doublet \citep{Massa2003,Bouret2005,Marcolino2009}. This line is usually 
unsaturated and, in principle, can serve as a robust mass-loss indicator.

\citet{Bouret2005} and \citet{Fullerton2006} found that mass-loss rates 
obtained from modeling P\,{\sc v} in O-star spectra are significantly 
lower than those measured from the emission H$\alpha$ line (see 
Section\,\ref{sec:clump}). \citet{Waldron2010} suggested that strong
emission  line radiation in the XUV band (between 54 to 124\,eV) near
the P\,{\sc v}  ionization edge   can significantly change the
ionization structure of the wind. This could  significantly reduce the
fractional  abundance of P\,{\sc v} and hence  skew the UV-line
diagnostics based on the  assumption that P\,{\sc v}  is the leading
ionization stage. They suggested that the  real mass-loss  rates  are
significantly higher, in agreement with the cool  optical depth 
estimated by \citet{WC2007} from the analysis of X-ray lines  in O-star
spectra. 

This idea was tested by NLTE stellar wind models that account for the 
influence of X-rays on ionization. PoWR models for a sample of 
supergiants showed that unrealistically high soft X-ray  fluxes are
required to significantly alter the  phosphorus ionization balance
(W.-R. Hamann, private communication).  \citet{Krt2012} studied the
influence of XUV radiation  on P\,{\sc v} ionization.  They confirmed
that  strong XUV radiation can decreases the P\,{\sc v} ionization
fraction.   However,  the XUV radiation would also  influence the
ionization  of elements that drive the wind, leading to a decrease of
the wind  terminal velocity which is not confirmed by observations.
Hence, it appears that albeit X-rays influence the  ionization structure
in the inner wind, they cannot be fully responsible  for the low
mass-loss rates derived from the analysis of the P\,{\sc v}  line.   

\citet{osk2007} and \citet[][and references therein]{Surlan2012} showed 
that the ``P\,{\sc  v} problem'' is resolved and consistent mass-loss
rates are obtained  from $\rho^2$  and $\rho$-dependent diagnostics 
when clumping is fully included  in the modeling of UV resonance lines. 
\citet{Owocki2008} and \citet{Sund2010}  used an alternative formalism
to describe specifically the effects of  a non-monotonic velocity and
coined the term  ``vorosity''. Using their 2D models, they found that for 
intermediate strong lines the velocity spans of the clumps are of central 
importance, albeit density structure and inter-clump medium also play a role. 

\citet{Surlan2013} performed advance 3D Monte-Carlo simulations of the 
formation of  resonance doublets in a clumped wind. In a detailed parameter 
study it was  demonstrated that ``vorosity'' has only a moderate effect. In 
any case, it seems that separating porosity and vorosity effects is rather 
artificial as both are playing role in establishing the line opacity.   
The concept of macroclumping (see Section\,\ref{sec:xprof}) accounts for both 
of these effects, and consistently includes them in the modeling 
\citep[e.g.][]{Surlan2013}.

\section{The ``weak wind problem''}
\label{sec:bw}

The  mass-loss rates of low luminosity ($\log{L_{\rm 
bol}/L_\odot}<5.2$)  main sequence OB-type stars, e.g.\ stars with
spectral  types O9V and later,  are orders of magnitude lower than
predicted by  ``standard''  mass-loss recipes 
\citep{Martins2005,Marcolino2009,Naj2011,osk2011}. This  is often
referred in the literature as the {\em weak wind problem}.

There are several hypotheses explaining why the mass-loss rates of OB
dwarfs empirically derived from UV diagnostics are very low. These
hypotheses can be divided into two broad groups: ({\it i}) some process
reduces  the wind driving force in OB dwarfs making the winds genuinely
weak,  and ({\it ii}) the large fraction of the wind is not visible in
the UV lines and  therefore cannot be measured  using UV diagnostics.
These theories are briefly  considered below. 

\subsection{Do X-rays reduce the wind driving in OB dwarfs and quench the 
wind?}

According to the  standard theory \citep[CAK][]{CAK1975}, stellar winds
are  radiatively driven. \citet{Drew1994} realized that the presence of
X-rays changes the  ionization structure in the inner part of the  wind.
The overionization  by X-rays may reduce the total radiative
acceleration and lead to a  lower mass loss. This suggestion was
investigated by  \citet{osk2011} using  X-ray observations of a sample
of B dwarfs. The  X-rays at the observed level  and temperature were
included in PoWR stellar atmosphere models, and  the wind radiative
force was computed. However, the models did not reveal a significant
inhibition of the wind driving power.    

On the other hand, \citet{Spring1992} questioned the  application of the
standard radiatively driven wind mechanism for B-type stars. They
showed that for low density winds the assumption  of a one-component
fluid is not valid. In such winds the metal ions loose their  dynamical
coupling to the ions of hydrogen and helium. The metal ions move  with
high velocities, while helium and hydrogen are not dragged along. The
collisionally induced momentum transfer is accompanied by frictional
heating, which dominates the energy balance. As a result,
\citet{Spring1992} predict electron temperatures of $T\lsim 1$\,MK in
the outer wind regions for stars with mass-loss rates of $\approx
10^{-8...9}$\,\myr. The subsequent analysis by \citet{Krt2000} and
by \citet{Owocki2002} favored even weaker winds with $\mdot\sim
10^{-11}$\myr\ for the frictional heating to occur. In this concept, 
the X-ray emission is not the cause for the small mass-loss rates 
empirically derived from UV and optical spectra, but its consequence. 

\citet{Krt2014} further elaborated the theory of mass loss from B 
stars. His theoretical mass-loss rates for main-sequence B-type stars
are  in good agreement with those empirically derived by
\citet{osk2011}. Furthermore,   it was shown that at solar abundance,
{\changed   a homogeneous line-driven wind is not possible for  stars 
with spectral types later than B5. This is in agreement with  the wind
limit found by \citet{Abbott1979}. }More recently, \citet{Mui2012} 
demonstrated that the well established Monte Carlo model  fails to
initiate radiatively driven winds for  main-sequence stars with
luminosities below $\log{(\Lbol/L_\odot)} \approx  5.2$. Consequently,
the standard mass-loss rate recipe \citep{Vink2001},  often used in the
literature, may severely overestimate the mass-loss rates of main
sequence B stars. 
%\citep[e.g.][]{Petit2013, Naze2014}

Another possible solution of the weak wind problem was suggested by
\citet{Martins2005}.  If present,  a sufficiently strong organized
magnetic field could confine  the stellar wind \citep{Babel1997}. The
wind streams collide in the equatorial  plane where strong shocks
produce relatively hard X-rays. The resulting  X-rays enhance the
ionization in the wind, destroy the lower ions  with many lines
responsible for wind driving, and thus quench the wind. 

Indeed, strong magnetic fields have been detected in some  OB dwarfs
such as HD\,54879 (O9.7V) and  HD\,57682 (O9IV)  \citep{Castro2015,
Grun2009}. The latter star has a weak wind and is a hard and  luminous
X-ray source ($\left<T\right>\approx 20$\,MK, $\log({L_{\rm X}/L_{\rm 
bol}}) \approx -6$). On the other hand, there are  O-type dwarfs where
the  presence of a strong magnetic field is ruled out, such as
$\zeta$\,Oph (O9.2IV)  and AE\,Aur (O9.5V)
\citep{Schnerr2008,Hubrigo2011}. These objects  have weak winds but a
normal level of soft  X-ray emission not sufficient to quench the
winds. 

Thus, it appears that at the moment there is no strong support for 
the idea of the wind quenching in OB dwarfs. 
 
\subsection{Is the large fraction of OB dwarf winds in the hot phase?}

The theory predicts rather low mass-loss rates for main-sequence
B-stars  \citet{bab1996}. If a fraction of a B star wind  gets heated,
e.g.\ in strong shocks,   the cooling time and length may be so long
that the bulk of wind material  may remain in the  hot phase
\citep[e.g.][]{Lucy1982,Abbott1989}. This  fraction  cannot be  detected
at optical and UV wavelengths, but is seen  in X-rays\footnote{Note
that  radio and IR observations may provide additional constraints, as
shown in  \citet[][p.29]{LC1999}}.

This idea is supported by X-ray observations. \citet[][and references 
therein]{Cas1994} noticed that the X-ray emission measure  ($EM_{\rm
hot}=\int_V n_{\rm e, hot}^2 {\rm d}V$) in near-main-sequence B stars 
is comparable or even larger than the cool wind emission measure, 
$EM_{\rm w}=\int_V n_{\rm e, cool}^2 {\rm d}V$ \citep[see also][]{Cohen2006}. 
This implies that the bulk of  stellar wind matter may be in a hot phase.
This hot fraction of the wind material is ''missed'' when mass-loss
rates are empirically deduced from the emission of the cool wind
observed as an excess in the infrared or radio continuum, or from
P\,Cygni line profiles  in the UV.  

\citet{Lucy2012} suggested that in late-type O dwarfs the shock-heating
of the ambient gas increases its temperature, and  the resulting
single-component flow with a temperature of a few MK coasts to high
velocities as a pure coronal wind. Hence, the volumetric roles of hot
and  cool gas are reversed in O dwarfs compared to O supergiants. The
bulk  of wind is hot, while  it contains only some cold clumps. 

X-ray observations of O-dwarfs seem to support this scenario. 
High-resolution X-ray spectra of $\zeta$\,Oph (O9.2IV),  $\mu$\,Col
(O9.5V), AE\,Aur (O9.5V), and $\sigma$\,Ori\,AB (O9.5V+B0.5V)  show
similar morphology \citep{osk2006, Skinner2008, Hue2012}.   The X-ray
luminosity of O dwarfs is a significant fraction of the  cool wind's 
mechanical luminosity.  For instance, the ratio between $L_{\rm X}$
and  $L_{\rm wind}=0.5\dot{M}v_\infty^2$ amounts to 0.1 in $\mu$ Col
and  0.7 in AE  Aur. There is only limited evidence for an absorption
of X-rays by the cool wind component, consistent with the above picture.
Analyses of line ratios  of He-like ions  (see Section\,\ref{sec:fir})
indicate that the hot matter is  present already in inner wind
regions where the wind velocity is low. The  X-ray emission lines are
broadened up to $v_\infty$, and  can be well  described as originating
from a plasma expanding with a  $\beta$-velocity law, i.e. the hot
plasma occupies a large span of  velocities. 

Thus, it seems that a new  picture of an OB dwarf stellar wind is 
emerging from theory and observation. The winds of O-dwarfs are
predominantly  in the hot phase, while the cool gas seen in the  optical
and UV  constitutes only a minor wind fraction.    Hence, the best 
observational window for studies of OB-dwarfs is provided by X-rays.  

\section{B-dwarfs in the X-ray light}
\label{sec:b}

As outlined above, X-rays provide an important window to study winds of
non-supergiant B-type stars that are hardly accessible in other bands
of the electromagnetic  spectrum. However, high quality X-ray data exist
for only a handful of non-supergiant B-type  stars, and these are
observationally biased to the closest and most  intriguing objects,
e.g.\ magnetic stars or binaries. 

A report on the diverse X-ray properties of some  magnetic B dwarfs can 
be found in  \citet[e.g.][]{osk2011,Naze2014}.  B-type binaries also
have diverse X-ray characteristics. $\theta$\,Car (B0.2V)  is an X-ray
luminous ($\sim 10^{31}$\,erg\,s$^{-1}$) binary with  soft X-ray
spectrum \citep{Hub2008, Naze2008}.  On the other hand, $\rho$\,Oph
(B2IV+B2V) has a quite hard variable X-ray spectrum \citep{Pil2014}. 
 
Perhaps the B-type star studied  best in X-ray is the early-type dwarf 
$\tau$\,Sco (B0.2V) \citep{Howk2000}. \citet{Mewe2003} obtained its
first  high-resolution X-ray spectrum. Its analysis  yielded
temperatures in the range of 1..20\,MK. The emission measure of the hot 
plasma is comparable to that of the cool wind. It was found that  the
spectral lines are not broadened above the instrumental profile. 
Correspondingly, the X-ray emitting plasma does not expand faster than
with $\sim  500$\,km\,s$^{-1}$, i.e. it is much slower than the cool 
wind that has a velocity of $v_\infty \sim 1000$\,km\,s$^{-1}$
\citep{osk2011}. From  the analysis of line ratios in the emission of
He-like ions it was shown that  the X-ray emission originates from
relatively close to the stellar surface  \citep[see
also][]{Cohen2003}.  

\citet{Donati2006} discovered a magnetic field on $\tau$\,Sco  with  a
complex  topology. This field should be capable to confine at  least
some portion of the stellar wind and, in principle, may help  to explain
the characteristic shapes of the UV lines. It was suggested  that the
X-ray emission  predominantly originates from closed magnetic  loops
that are distributed  non-uniformly over the stellar surface. Because of
rotation, the  active regions cross our line of sight. If produced in
these active regions, the rotational  modulations of X-rays should be up
to 40\%\ \citep{Donati2006}.   Surprisingly,  monitoring of $\tau$ Sco
over its rotation period  with the {\em Suzaku} X-ray observatory did
not reveal the predicted  variability, indicating that the X-rays are
likely produced in  small-scale magnetic loops distributed across the
whole stellar surface \citep{Ignace2010}.

Recently, it was confirmed that a whole group of stars that have  UV
spectral features  similar to $\tau$\,Sco also host magnetic fields; 
these stars, HD\,66665 and  HD\,63425, were dubbed $\tau$ Sco-analogs
\citep{Petit2011}.  \citet{Ignaceb2013} investigated their X-ray
properties, and found them similar to those of $\tau$ Sco. Thus, it
appears that $\tau$\,Sco  analogs have similar magnetic, stellar wind,
and coronal properties. 

The sample of non-magnetic B-dwarfs studied in X-rays is rather small.
The  star $\sigma$\,Sgr (B2.5V) was only marginally detected by  \xmm\
in a 10\,ks exposure. Its X-ray luminosity  $\Lx\approx  4\times
10^{27}$\,erg\,s$^{-1}$ ($\log{\Lx/\Lbol} \approx -9.4$)  (Oskinova et
al.\,in prep) is  much lower than previously reported  from  {\em 
Rosat} observations.  Possibly, the latter measurement was contaminated
by a close and X-ray variable   low-mass coronal companion
\citep{Gul2013}. Active low-mass  stars may often outshine their more
massive companions  in the X-ray light  \citep[e.g.][]{Evans2011}. 

\subsection{Pulsating $\beta$ Cep-type stars} 
\label{sec:bcep}

Young hydrogen-burning B0--B2 type stars pulsate with periods of a few 
hours. These stars are called $\beta$ Cep-type variables. The
oscillations  are driven by changes of the opacity inside the star
during  the pulsation cycle  \citep[``$\kappa$-mechanism'',][]{dz1993}.
Like  other early B-type stars, $\beta$ Cep-type variables are X-ray
sources.  

Besides radiatively driven stellar wind and magnetic effects, stellar 
pulsations may play a role in powering the X-ray emission from B stars. 
E.g.,\ \citet{Neilson2008} and \citet{Engle2014} demonstrated that 
pulsations may heat the outer atmosphere of classical  Cepheids   and
power X-ray  emission even in these cool stars.

{\changed  Observations with the {\em EUVE} satellite showed that
$\beta$\,CMa (B1II/III) and  $\epsilon$ CMa (B1.5II) have an order of
magnitude excess in their  extreme UV (EUV) spectra relative to the
predictions of stellar atmosphere  models.  } It was suggested that the
deposition of mechanical energy from stellar  pulsations can heat the
inner wind regions causing the  observed excess \citep{cas1996}.  W.-R.
Hamann  (priv. communication) showed  that the  observed extreme UV
excess in $\beta$ and $\epsilon$\,CMa could  be reproduced by
accounting  for a continuous temperature distribution of the X-ray
emitting plasma through shock heating and the subsequent cooling
sequence.  Interestingly,  \citet{Fos2015} detected weak  magnetic
fields on   $\beta$\,CMa (B1II/III) and $\epsilon$\,CMa (B1.5II).  

If stellar oscillations are somehow responsible for the X-ray emission 
from $\beta$\,Cep variables, one may hope to find correlations between
the pulsational behavior and  the X-ray properties.  An initial X-ray
survey of six $\beta$\,Cep-type stars was performed with  the {\em
Einstein} observatory \citep{agr1984}, but no correlations were  found
between X-ray  and pulsational, rotational, or binary properties. 
Negative results were also obtained in the study of four \bcep\ stars
observed  with  {\em Rosat} \citep{Cas1994}. \citet{osk2011} searched
for correlations  between pulsational and X-ray  properties in a small
sample of \bcep-variables  with well known pulsational behavior and
existing X-ray observations, but did  not find any. Timing analyses of
$\beta$ Cen and $\beta$ Cep using  very  high quality X-ray light curves
obtained with \xmm\ and {\em Chandra} did  not reveal any X-ray
oscillations \citep{raas2005,fav2009}. Both these stars are  magnetic
\citep{Donati2001,schnerr2006,alecian2011}. A report on  detected X-ray
pulsations of $\beta$ Cru \citep{Cohen2006} could not be  confirmed by
an independent study \citep{osk2015}. 

Stable, coherent X-ray pulsations in phase with the optical light-curve 
but even larger amplitude were firmly identified so far only in one B
type star -- $\xi^1$\,CMa  (B0.7IV) \citep{osk2014}. This star has the
strongest  magnetic field among all  $\beta$\,Cep-type variables with
available  measurements \citep{Hub2006}. The  phase resolved X-ray
spectroscopy  using  \xmm\ data revealed that the X-rays originate from
close to the  stellar photosphere, questioning the  prediction of the
magnetically confine wind shock model \citep[][for  the latest
development of the model  see the review of ud-Doula \& Naz\'e in this 
volume]{Babel1997}. 

\section{X-ray variability as a diagnostic of inhomogeneous stellar winds}
\label{sec:xvar}

Obtaining X-ray light curves and establishing parameters of X-ray 
variability provides a valuable tool to gain insight into the stellar 
wind structure.

There are numerous observational evidences supported by theoretical 
considerations for stellar winds being structured and inhomogeneous 
\citep[e.g.][]{Hamann2008}.  Among the first clear demonstrations of
wind inhomogeneity  was the detection of stochastic variability in  the
He\,{\sc ii}\,$\lambda$4686\,\AA\ emission line in the spectrum of an O
supergiant \citep{Eversberg1998}, explained  by  clump propagation.
\citet{Markova2005} investigated the line-profile variability of 
H$\alpha$ for a large sample of O-type supergiants and concluded that
the  observed variability can be explained by a structured wind
consisting of  shells fragments. Using spectral diagnostics
\citet{Prinja2010} showed  that the winds of B supergiants are clumped.
In a recent study  \citet{Martins2015} found that spectral lines formed
in the winds of all OB  supergiants in their sample are variable on
various time scales.  \citet{Lepine1999, Lepine2008} monitored the
line-profile variations in a  sample of WR and O stars and explained
their observations using a  phenomenological model  that depicts winds
as being made up of a large  number of randomly  distributed, radially
propagating clumps.  

Convincing evidence of wind clumping is provided by high-mass X-ray
binaries (HMXBs). In some of these systems, a neutron star (NS) is in a
close orbit deeply inside the stellar wind of an OB star. Accretion from
the clumped stellar wind onto the NS powers strongly  variable X-ray
emission \citep[e.g.][]{Bozzo2011,osk2012,Martinez2014}. 
\citet{vdMeer2005} studied the X-ray light curve and spectra  of 4U
1700-37 and concluded that the feeding of the NS by a strongly clumped 
stellar wind is consistent with the observed stochastic variability. 
Further evidence of donor wind clumping comes from the  analysis of the
X-ray spectra.  \citet{Schulz2002,Gim2015} reviewed the  spectroscopic
results obtained with X-ray  observatories for wind-fed HMXBs.  They
explained the observed spectra  as originating in  a clumped stellar
wind, where cool dense clumps are embedded in rarefied photoionized gas.
\citet{Tor2015} used fluorescence lines as a tracer of wind clumps and 
determined that these clumps must be  present close to the photosphere
in case of a B-type supergiant. 

One can expect that, similar to the variability observed in optical and UV 
light,  the X-ray emission from OB supergiants should also be variable.
Indeed, hydrodynamic models predict strong stochastic X-ray variability with 
very large amplitude, albeit this may be an artifact of the 
1-D geometry of hydrodynamic simulations  \citep{felda1997,feld1997}. 

Yet, strong stochastic X-ray variability is not observed. 
\citet{osk2001}   developed a phenomenological wind model and showed
that the character of  X-ray variability depends on the number of shocks
as well as on the cool wind  opacity. Consequently, the observations of
X-ray variability can provide   important information on the structure
of stellar wind.   

Given its large diagnostic potential, there were numerous attempts to
measure the X-ray variability of O stars. \citet{Berghoefer1996} used 
the X-ray telescope {\em Rosat} (0.9-2.4\,keV) to monitor  $\zeta$\,Pup
over 11 days totaling 56\,ks observing time. Contemporary with X-rays
the variability of H$\alpha$ was also monitored.  They reported a
16.667\,h modulation in the H$\alpha$ line correlated with  modulations
in the X-rays. The amplitude of the X-ray variability was found to  be
$\pm 6$\%. These results were interpreted as an evidence for periodic 
variations in the wind density.  

\citet{osk2001} analyzed archival X-ray observations of $\zeta$\,Pup
and  $\zeta$\,Oph made with the X-ray telescope ASCA (0.5-10 keV). ASCA 
observed $\zeta$\,Pup continuously for 28.4\,h. The timing analysis of
the data  did not reveal modulations as those reported by
\citet{Berghoefer1996}. On the other  hand, modulations on the time
scale of days were found in the  X-ray emission  of $\zeta$\,Oph.  

\citet{Massa2014} analyzed X-ray observations 
of $\xi$\,Per  (O7.5III) obtained with the {\em Chandra}
X-ray telescope and contemporaneous  H$\alpha$ observations. The 
X-ray flux was found to vary by $\sim 15$\%, but  not in phase with 
the H$\alpha$ variability. The observations were not  long enough
to establish periodicity.

Among the longest time intervals covered by X-ray observations of  a
particular star is the monitoring of $\zeta$ Pup by \xmm. About  20
exposures of various duration were obtained during  a decade.  Timing
analysis of these observations did not reveal the period  reported by
\citet{Berghoefer1996}. However, variations with amplitude of $\sim
15$\%\ on a time scale longer than  1\,d were found, while no coherent 
periodicity was detected \citep{naze2013}.

\citet{Rauwb2015} obtained X-ray and coordinated  H$\alpha$
observations  of $\lambda$ Cep (O6.5I(n)).  They found that the X-ray
flux varies by $\sim 10$\,\%\ on  timescales of days and might be
modulated by the same period as the H$\alpha$ emission, but  shifted in
phase.  The analysis of archival \xmm\ observations of $\zeta$ Ori  
shows that the X-ray variability of this star has similar properties  to
that of $\lambda$ Cep and $\zeta$ Pup. X-ray variability with analogous 
character was found in $\delta$ Ori (O9.5II+B1V) from {\em Chandra} 
observations, where periodic X-ray fluctuations (not associated  with
the binary period) were identified \citep{Nichols2015}. Even WR stars
with  very strong winds show modulations in their X-ray  emission on the
rotation timescale \citep{Ignace2013}. 

Thus, solid evidence of slow X-ray variability of O-type star is
accumulating.  The new high-quality data suggest an association between 
X-ray emission and large scale structures in the stellar wind.

It seems, that the X-ray variability is closely related  to the coherent
and periodic  variability  that is commonly observed in the UV spectral
lines of OB supergiants \citep[e.g.][]{Kaper1999, Massa2015}. The
latter  is explained  by the existence of corotating interaction
regions  (CIRs) in stellar winds, which can also play a role in the 
generation of X-rays \citep{Mullan1984}. \citet{Cranmer1996} showed that
CIRs could result from bright stellar spots.  \citet{Ram2014} detected
small corotating  bright spots on $\xi$\,Per\ and suggested  that they
are  generated via a breakout of a global magnetic field generated by 
subsurface convection. The CIRs may also be triggered by the
(non)radial  pulsations  of the stellar surface \citep{Lobel2008}.

\section{Modeling of X-ray emission line profiles}
\label{sec:xprof}

In this section, the diagnostic power of X-ray emission lines to probe
stellar  winds is considered. Presently, lines are resolved in the X-ray
spectra of  nearly all types of massive stars. Line shapes and fluxes 
provide valuable  information about the motion and geometrical
distribution of the hot plasma,  and allow to probe stellar wind
opacity. In general, the X-ray emission  lines are broad, indicating
that the hot plasma is expanding with high velocities,  comparable to
the wind velocity obtained from UV line measurements 
\citep[e.g][]{WC2007}. Moreover, the shape of the emission lines in O
star spectra shows that the hot plasma  is spread over a large range in
the velocity space \citep[e.g.][]{Cor2015} 

The shape of an X-ray emission line can be calculated from a solution of
radiative transfer equation describing the emission from expanding
optically thin shell and its absorption in the smooth cool stellar wind 
with high continuum opacity  \citep{Macfar1991}.

The optical depth along the radial line of sight is given by
\begin{equation}
 \tau_\lambda=\int_{r_0}^\infty \chi_\lambda(r){\rm d}r,
\label{eq:taur} 
\end{equation}
where $r_0$ is expressed in the units $R_\ast$ and denotes the onset of X-ray 
emission, and the atomic opacity 
\begin{equation}
 \chi_\lambda = \rho_{\rm w} \kappa_\lambda 
\end{equation}
is the product of the mass 
absorption coefficient $\kappa_{\lambda}$ [cm$^2$\,g$^{-1}$] and the 
density of the cool wind ($\rho_{\rm w}$). The latter obeys the continuity 
equation 
\begin{equation}
 \dot{M}=4\pi\rho_{\rm w}(r) v(r) r^2 R_\ast^2,
\label{eq:dotm}
 \end{equation}
where $r$ is the radial coordinate in the units of $R_\ast$, and $v(r)$ is the 
velocity law. Commonly, a $\beta$-velocity law is adopted 

\begin{equation}
v(r)=v_\infty\left(1-\frac{1}{r}\right)^\beta. 
 \label{eq:bv}
\end{equation}

%HERE 01 March 2016

\citet{Macfar1991} applied their formalism to the very soft X-ray (XUV) 
radiation of B supergiants. They reasoned  that because the stellar wind is 
quite opaque at these wavelengths, the observed radiation must originate 
from outer wind regions, where the wind ionization  is constant. In this case  
one can neglect the radial dependence of the mass absorption coefficient 
$\kappa_{\lambda}$ and Eq.\,(\ref{eq:taur}) can be simplified to
\begin{equation}
 \tau_\lambda=\tau_\ast\int_{R_0}^\infty 
\left(1-\frac{r_0}{r}\right)^{-\beta}r^{-2}{\rm d}r,
\label{eq:taul}
 \end{equation}
where  
\begin{equation} 
\tau_\ast=\frac{\kappa_\lambda \dot{M}}{4\pi v_\infty R_\ast}.  
\label{eq:t0}  
\end{equation}  
\citet{Macfar1991} pointed out that the line shape is largely determined  by 
the parameter $\tau_\ast$. When $\tau_\ast$ is small, the line is broad and 
has a box-like shape. For  stronger wind absorption, the line becomes more 
skewed  \citep[see Figure\,7  in][]{Macfar1991}, because the photons in the
red-shifted part of the line are more attenuated than in the blue-shifted 
part. \citet{Macfar1991} showed that evaluating the line shape can be used to 
determine $\tau_\ast$, and thus probe the wind opacity.

The opacity for X-rays is largely determined by the K-shell absorption 
in metal ions. The K-shell opacities, and consequently the coefficient 
$\kappa_\lambda$, vary with wavelength with a power between 2 and 3 
\citep{ver1995}. Therefore, $\tau_\ast$ should change by orders of
magnitude over the  X-ray band ($\sim 1 - 30$\,\AA). Thus, the shape of 
emission lines should be different at longer and at shorter wavelengths.

The \citet{Macfar1991} formalism turned out to be extremely useful in 
describing the X-ray emission line profiles when the first
high-resolution X-ray spectra of OB stars were obtained with \xmm\ and
\cxo.  Contrary to the predictions, the observed X-ray lines showed
only  little  asymmetry, and were found to have similar shapes at
different  wavelengths \citep{wc2001,Kahn2001,cas2001}. 

\citet{wc2001} considered emission from spherically symmetric shocks equally
distributed between 0.4\vinf\ and 0.97\vinf\, with temperatures ranging
from 2 to 10\,MK. They found that the mass-loss  rate of the O-type
supergiant $\zeta$\,Ori must be lower than traditionally  assumed in
order to explain the observed nearly symmetric X-ray emission line  profiles.

\citet{oc2001} updated the formalism from \citet{Macfar1991}  by 
including a $\beta$-velocity law and allowing for a radial dependence 
of the hot gas filling factor. Independently, \citet{Ignace2001}
considered expansion with a constant  velocity but included a radial
dependence of the filling factor.  

\subsection{X-ray line profiles from clumped stellar winds}
\label{sec:clump}

Stellar wind clumping was included in the line formation formalism by 
\citet{feld2003}.  The stellar wind structure affects the propagation of
X-rays and the emergent   X-ray line profile. In a first approximation 
(``microclumping''), clumping is taken into account in up-to-date
stellar wind codes  \citep{hil1991,hk1998,puls2006}.  This approximation
assumes that all clumps  are optically thin. The interclump medium is
void or filled with  tenuous matter \citep[][]{Zs2008}. In this
so-called microclumping approximation  the radiative transfer is not
affected by the clumping,   which makes this approximation very
convenient. One usually introduces a  parameter $D$ to describe  the
factor by which the density in the clumps is enhanced compared to a
homogeneous wind with the same mass-loss rate. If the interclumped 
medium is assumed to be void, the clump volume filling  factor is
$f_{\rm V} = D^{-1}$.  The most important consequence of microclumping
is the reduction of mass-loss rates empirically derived from diagnostics
that depend on  density squared (such as H$\alpha$ emission fed by the
recombination  cascade,  and free-free radio emission) by  a factor
$\sqrt{D}$ \citep[e.g.][]{hk1998,Bouret2005,Fullerton2006}. 

%============================ Figure  ===============================
\begin{figure}[t]
\centering
\includegraphics[width=\columnwidth]{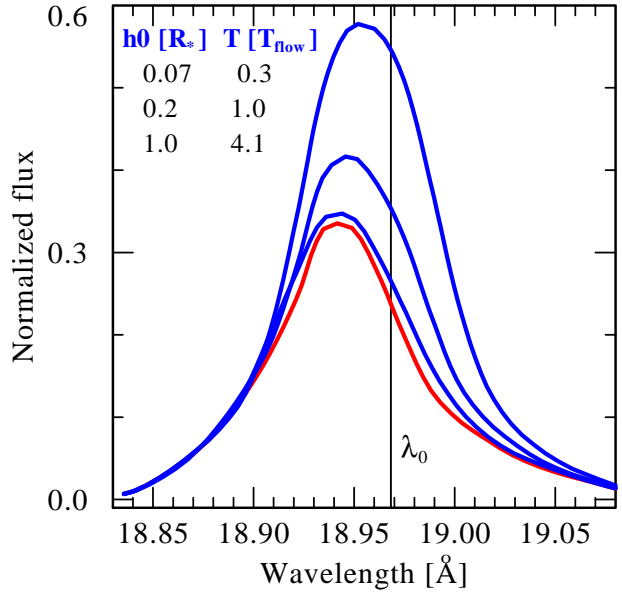}
\caption{The effect of clumping parameter $T$ (or
$h(R_0)$) on model lines assuming anisotropic opacity with clumps flattened 
in radial direction, e.g. broken shell fragments.  Except of the 
clumping parameter $T$ ($h(R_0)$), model parameters are for the star 
$\delta$\,Ori and are the same for each model 
line. The red line is for smooth wind, while blue lines are for clumped 
wind models with porosity length $h(R_0)<1$. The line computed with the 
smallest porosity length is similar to the smooth wind model.}
  \label{fig:pors}
\end{figure}
%=======================================================================

Microclumping is a stringent approximation, and one  has to  be aware of
its limitations. Waiving the microclumping approximation in  modeling is
called ``macroclumping''. Macroclumping  does not impose any 
restrictions on a clump's optical  depth. The same clump could be
optically  thin  at some wavelength and optically  thick at another. 
This  has implications for  the analysis of X-ray spectra from OB
supergiants -- the lines at shorter  wavelengths may not be affected,
while clumping  effects become more pronounced for lines at longer
wavelengths  \citep{osk2006}. \citet{Ignace2016} studied the influence
of macroclumping  on the free-free  spectral energy distribution of
ionized winds, with applications to the  formation of radio spectra and
to mass-loss rate diagnostics. 

A general theory for the transfer of X-ray radiation in clumped stellar
winds, including limiting cases of optically thin and fully opaque clumps, as 
well as intermediate cases,  was developed in a series of papers by
\citet{feld2003,osk2004,osk2006}.  An analytic description for the
effective opacity in  isotropic  and anisotropic cases was found. 
(Figure\,\ref{fig:pors} demonstrates lines emerging from winds with 
anisotropic opacity.) It was pointed out that in case of  isotropic
opacity (e.g.\ spherical clumps) the line {\em profiles} are  identical
to those emerging from a smooth wind. On the other hand, in case of  
anisotropic opacity, the  line profiles are characteristically
different. The  analytic solutions were verified by  2.5-D Monte-Carlo
simulations   \citep{osk2004}, and reproduced by, e.g.,\
\citet{sund2012} using their  alternative approach. Detailed
considerations of X-ray line formation in  stellar winds are given in
the review by R.\,Ignace (this volume).   

To specify clumping properties, \citet{osk2001,osk2004} used 
a ``fragmentation frequency'' parameter $n$.  One can envision this as the
number of clumps that are ``launched'' from  some initial radius per dynamical 
time  $T_{\rm fl}=R_\ast/v_\infty$. Assuming 
that  clumps do not merge and do not disappear  until some outer 
boundary, the  fragmentation frequency $n$ is radius independent. 
The inverse of the fragmentation  frequency, $T=1/n$, is the average 
time between two successive clumps passing  through some point in the
wind.  

\citet{oc2006} and \citet{leu2013} used the ``porosity'' formalism to
model X-ray  emission lines emerging from clumped winds. 
\citet{Herve2012} provided a  careful comparison of the macroclumping
and porosity formalisms and  concluded that they are ``essentially
equivalent to first order''. 

\citet{Owocki2004} and \citet{oc2006} use the porosity  length $h$ to 
parametrize the clumping properties. This porosity length is the ratio
of the volume  per blob to the projected surface area of the blob, $h =
L^3/l^2$.  Equivalently, it is the blob size divided by its volume
filling factor, $h =  l/(l/L)^3$ \citep{Owocki2004}. Usually, it is
assumed that the porosity  growth with the radial coordinate   and
reaches  its maximum value at $v_\infty$. \citet{oc2006} argued that
porosity  effects become important only for unrealistically large
porosity length $h \gg  1$. However, \citet{osk2006} showed that 
because of the strong wavelength dependence of the wind opacity,
porosity effects are also strongly wavelengths dependent. As 
illustrated in Fig\,\ref{fig:dege}, realistically small porosity 
lengths are sufficient to influence X-ray lines.  

In the hydrodynamic models  by \citet{feld1997} the X-rays are 
generated when a fast parcel of gas rams into a slower moving dense 
shell. \citet{feld2003} solved  the X-ray radiative transfer in such
situation, and showed that the emerging  line profiles have a strong
characteristic dip at the line center. In the hydrodynamic  models by 
\citet{Ignace2010} the X-rays are emitted from  bow shocks created by
dense clumps.  These models also predict a  dip in the center of the
emergent line 

\begin{figure}[t]
\centering
\includegraphics[width=\columnwidth]{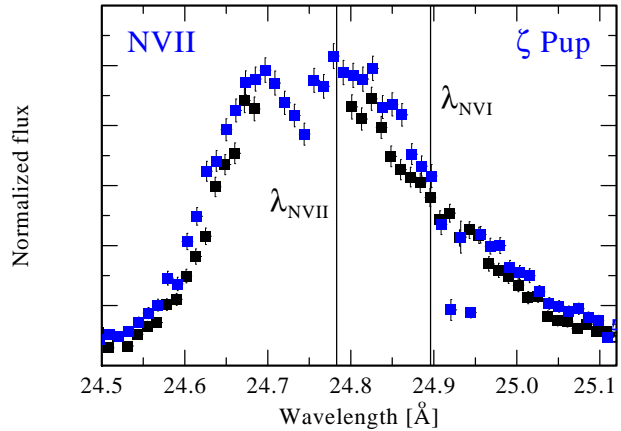}
\caption{The N\,{\sc vii} $\lambda$\,24.758\,\AA\ and N\,{\sc vi}
$\lambda$\,24.898\,\AA\ blend in the \xmm\ RGS1 (blue squares) and RGS2 
(black squares) spectra  of \zpup. The central wavelengths of the lines are 
indicated.  The absorption dip at $\approx 24.75$\,\AA\ is characteristic
for non-negligible line optical depth. Figure produced from archival data.
\label{fig:nvii}}
\end{figure}

The quality of available X-ray spectra allows to study fine details  of
X-ray line profiles. It appears  that a central dip is indeed observed
in some  strong X-ray lines  (see e.g.\ Fig.\,\ref{fig:nvii}). 
Alternative explanations of the central  dip may include resonance
scattering  \citep{oska2006}. Profile shapes for  optically thick X-ray 
emission lines from stellar winds were modeled by \citet{Ignace2002}.

%------------------------------------------------
%
%
%------------------------------------------------------------------

\section{Using X-ray spectra to probe stellar wind opacity and mass-loss}
\label{sec:obi}

The winds of OB supergiants are strong  and efficient in absorbing
X-rays.  Therefore, the  X-ray spectra could be exploited to provide 
useful  information on wind column density, abundances, and structure. 

\citet{Waldron1984} used calculated wind opacities and demonstrated that
the  analysis of X-ray spectra  can be used to probe the latter.  To
describe the  propagation of broad-band X-ray radiation in stellar
winds,  \citet{oc1999},  \citet{Ignace1999}, \citet{Ignace2000} adopted
the {\em exospheric  approximation} in which all of the emission from
above the sphere with the  radius where the optical depth is unity
escapes, while all emission from below  this radius is absorbed.
{\changed This is a reasonable approximation because   the
``contribution function'' to a line  peaks at the radius where wind
optical  depth is close to unity.} A more precise solution was presented
by   \citet{leu2010}. 

\citet{hil1993} computed the cool wind opacity of $\zeta$\,Pup  assuming
a smooth wind and applied their model to the low resolution  {\em Rosat}
X-ray spectrum. It was found that the high opacity of the stellar  wind
should completely block the soft X-rays ($<0.5$\,keV).  However, since
such  soft X-rays are observed, it was concluded that a significant
fraction of the hot  plasma is located far out in the wind, at distances
exceeding 100\,\Rstar. 

However, the analysis of high-resolution X-ray spectra of O stars (see  
Section\,\ref{sec:fir}) revealed that the X-rays  are  produced close to
the photosphere. The question how those X-rays  could pass through the
overlaying stellar wind without being  completely absorbed has been
subject of intensive studies over the past fifteen years.

At least two explanations for the effectively low wind opacity for
X-rays were  suggested: porosity  of a clumped stellar wind
\citep{feld2003}  and  lower mass-loss rates  than commonly adopted
\citep{Kramer2003}.

Besides these explanations, a non-symmetric two-component wind
structure  was  considered by \citet{Mullan2006}. They suggested a 
scenario where one wind component  emerges from magnetically active
polar regions, while the second, radiatively driven component,
originates  in a broad range of latitudes centered  on the equator. Yet
another idea was suggested by  \citet{Pollock2007} who proposed that
X-rays  originate in the wind's terminal velocity regime in
collisionless shocks controlled by magnetic fields rather than in
cooling shocks in the acceleration zone. 

\citet{Zhekov2007} analyzed the spectra of 15 massive OB stars and found
that their sample stars fall into two groups: stars with relatively
soft and with relatively hard X-ray spectra.
\citet{Zhekov2007} cautioned that the origin of X-ray emission is likely
different between these groups. While X-rays may originate from 
small-scale wind shocks in stars with softer spectra, magnetically
confined wind shocks are likely to be responsible for the X-ray emission
of the other group.

\subsection{Measuring mass-loss rates using X-ray line profiles?}
\label{sec:mdot}

Recently, it was suggested that a star's mass-loss rate can be 
determined by fitting the simple quantitative model  of \citet{oc2001}
to each emission line in its ({\rm X-ray})  spectrum  and then
analysing  the ensemble of line optical depth  \citep{coh2011}.  If
valid, this method would provide an interesting diagnostic of stellar
mass-loss rates. However, the method is based on the assumption that
the   absolute values for the wind opacity derived by \citet{Mac1994} 
are {\em universal} for all O-type  stars \citep{coh2014}. 

The universality of wind opacity is a stringent assumption. It assumes 
that all O-type stars have the same -- solar -- chemical composition,
and the  same ionization stratification of their winds. Yet, detail
non-LTE calculations  of wind opacities show strong departures from
these universal values. For   instance, \citet{Herve2012} computed
detailed stellar  atmosphere models of $\zeta$\,Pup. At
$\lambda=19$\,\AA\  their models predict a factor of $\sim 2$ larger
$\kappa_\lambda$  than  the universal value. These uncertainty would
directly propagate to the  estimated  mass-loss rate.  Similarly,
\citet{Car2016} recently discussed in detail  the behavior of the
mass-absorption coefficient in O star winds. Their detailed  models of
wind opacites do not comply with a generic universal values of 
$\kappa_\lambda$ for all O stars.

Wind clumping and radial dependence of filling factor, which are
neglected  in \citet{coh2014}'s method, also may skew the mass-loss
diagnostics. By fitting observed  X-ray line profiles, one  principally
cannot distinguish between clumped models with isotropic opacity  and
smooth wind models  (see Section\,\ref{sec:clump}). Therefore, the 
good  fits of observed line profiles obtained with  smooth wind models
indicate at best, that the stellar wind clumps are more or less
isotropic in shape (i.e.\  spherical or randomly oriented). 

%============================ Figure  ===============================
\begin{figure}[t]
%\centering
\includegraphics[width=0.8\columnwidth]{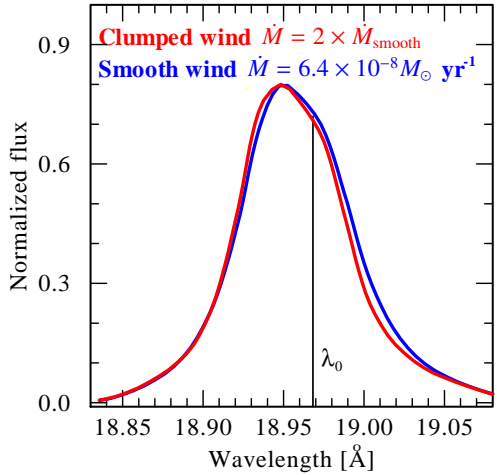}
\caption{X-ray model line profiles. 
The vertical line indicates the central wavelength of the O\,{\sc 
viii} L$\alpha$ line. Two model lines 
computed with identical parameters except of mass-loss rate and
clumping are shown. The universal  mass-absorption coefficient $\kappa$ is 
adopted from  \citet{coh2014}. The blue line is for a  
smooth wind  with an onset of X-ray emission at $R_0=1.33\,R_\ast$ 
and $\dot{M}=6.3\times 10^{-8} M_\odot$\,yr$^{-1}$. The red line 
has all the same parameters except that macroclumping with 
isotropic opacity (spherical clumps) is included and the mass-loss rate is 
two times larger than for the smooth wind model. The clumped wind model 
corresponds to a porosity length of $h(R_0)=0.1$.  The 
model lines are convolved with  {\em Chandra}'s MEG spectral 
resolution. 
%These two model profiles are
%practically indistinguishable, hence demonstrating the strong degeneracy 
%between clumping properties and mass-loss rate.
 }
  \label{fig:dege}
\end{figure}
%=======================================================================

To test how much clumping affects the mass-loss estimates, we computed a
grid  of model lines with various clumping parameters, ranging from very
strong clumping to smooth winds. As an example,  Figure\,\ref{fig:dege}
shows two similar model lines, one from emerging from a  smooth, and
another one from a clumped wind.  The figure illustrates that  fitting
of the line profile, in principle,  cannot provide clumping independent
mass-loss rate diagnostics. Even a small  porosity length $h$ or high
fragmentation frequency $n$ (see Section  \ref{sec:clump}) gives a
factor of two different mass-loss rate  when the latter is evaluated
from fitting the line shapes. 

Finally, the detailed stellar wind modeling shows that the observed
X-ray  spectra cannot be reproduced consistently with UV and optical
spectra unless  either macroclumping or a radial dependence of the hot
gas filling factor is included in the  model atmospheres
\citep{Herve2013,Shenar2015, Puebla2016}.

Thus, measuring stellar mass-loss rate by fitting simplified models to 
the observed X-ray spectra  seems to suffer from a parameter degeneracy (it 
represents an ill-posed inverse problem). Mass-loss rates cannot be reliably 
derived from the ensemble of X-ray line profiles without additional information on 
ionization structure, abundances, clumping, and velocity law in stellar wind.

\section{Combining X-ray spectroscopy with multiwavelength analysis to 
empirically constrain stellar wind properties.}
\label{sec:nlte}

Over the last decade X-ray spectroscopy was successfully combined  
with traditional methods  of spectroscopic analysis. The conditions in
stellar winds strongly depart from LTE,  and sophisticated codes have 
been developed to tackle the problem of spectrum  formation in this 
situation. The challenge is to include an X-ray radiation field in these 
stellar atmosphere models and to obtain a synthetic X-ray spectrum  {\em
consistently} with UV and optical spectra that can be compared
with observations.     

The PoWR non-LTE stellar atmosphere model accounts for the effect of the
X-ray  field on the cool wind. \citet{Shenar2015} used PoWR models  to
study X-ray, optical, and UV spectra of the O star $\delta$\,Ori\,Aa
(O9.5II).  They adopted a radius dependent clump volume filling factor,
with the maximum value $f_\infty=0.1$ reached at $\sim 10\,R_\ast$ as
motivated by hydrodynamic wind simulations. Using macroclumping  it was
possible to reproduce the UV and optical spectra as well as the X-ray
emission lines consistently.  The stellar mass-loss was found to be in
good agreement with  hydrodynamical predictions \citep{Vink2001}.

\citet{Herve2012,Herve2013} and \citet{Rauwb2015} included the X-ray 
radiation in the non-LTE  CMFGEN stellar atmosphere models to obtain
synthetic  high resolution X-ray spectrum. The best agreement between
the model  and the observations was achieved with a non-monotonic radial
distribution of  the X-ray filling factors and the location of the X-ray
plasma in $\lambda$\,Cep  (O6.5I) between $1.1\,R_\ast$ and
$2.5\,R_\ast$, and in $\zeta$\,Pup between  $1.5\,R_\ast$ and the outer
wind regions. Global fits of the X-ray spectra  yielded mass-loss rates
in very good agreement with previous determinations based on optical and
UV spectra.

A consistent analysis of the X-ray, UV, and optical data of the B0Ia 
supergiant $\epsilon$\,Ori was performed by \citet{Puebla2016} using 
the CMFGEN code. A radial dependence of the X-ray filling factor  was
used as  free parameter to bring the model spectrum in accordance with 
observations. (Note, however, that \citet{coh2010} tested models with 
varying filling factors by fitting them to observed X-ray  emission
lines and concluded that a radial dependence of the filling factor can
be neglected  since this does not improve the line fits.) Macroclumping 
was not  included in the models, but very strong microclumping with
$f_\infty < 0.01$ was required to reproduce the UV spectrum of Si\,{\sc iv}
line.  Under these assumptions, the mass-loss rate would be  
$\dot{M}<1\times 10^{-7}\,M_\odot$\,yr$^{-1}$. 

\section{X-ray diagnostics of massive stars in the latest stages of their 
evolution: RSG, WR, and LBV stars}
\label{sec:wr}

In its late evolutionary stage, a massive star could become a red
supergiant  (RSG), luminous blue variable (LBV), or/and a Wolf-Rayet
(WR) star. The exact  evolutionary path depends on various parameters,
such as initial mass, rotation,  metallicity, and binarity 
\citep[e.g.][]{Hamann2006,Langer2012,Sander2012}.

Red supergiants are X-ray dark. Searches with \xmm\ for X-ray emission 
possibly associated with surface magnetic activity or shocks in the RSG
winds  resulted in null detections \citep[][and references
therein]{Montez2015}. The  winds of RSGs are strong, therefore the null
result may not be surprising,  even if any X-ray emission were generated
in lower wind layers, it would be  absorbed in the wind. 3D
hydrodynamic  models predict that in binary systems, such as e.g.\ 
Antares  ($\alpha$ Sco, M1\,Ib + B2.5\,V), the blue companion moving
through the outer  RSG wind produces a photoionized cavity and a wake 
\citep{Braun2012}. In principle, one might speculate that X-rays are 
generated either in the wind-wind collision zone or in the
photoionization  wake. However, Antares is X-ray dark.

The enigmatic L-type supergiant V838\,Mon may be a merger product
between a  massive B-type star and an intermediate mass star
\citep{Tylenda2006}. It is  theoretically expected that merger products
may have strong magnetic fields  \citep{Tutukov2010}. Therefore, the
merger hypothesis could be potentially  tested by X-ray observations of
the post-merger object. The post outburst  observations of V838\,Mon did
not reveal X-ray emission. However, later on a new X-ray source was
found a few arcsec away from V838\,Mon. One of the  possible
interpretations of this X-ray source is the interaction between the 
V838\,Mon ejecta and the B3\,V companion \citep{Anton2010}.

Interestingly, it was recently suggested that a merger event triggered
the  the great eruption of  $\eta$ Carinae \citep{PZ2015}. This  LBV is
one of the most massive stars in the Galaxy and a prodigious X-ray
source  \citep{Seward1979}. The X-rays in $\eta$ Car are produced in a 
collision of the LBV wind with the wind of its early-type companion 
\citep[e.g.][]{Dam2000,Cor2001,Ham2007,Parkin2009}. As in other
colliding wind  systems, X-rays provide important information about
orbital, stellar, and wind  parameters (see review by Rauw \& Naz\'e in
this volume). 

Another X-ray bright  LBV star is HD\,5980 in the Small Magellanic
Cloud. The X-ray emission from  this system is also powered by the
collision between stellar winds in a  multiple  system \citep{naze2007}.
In general, it seems that all LBV stars detected in   X-rays are
multiple systems, where collisions between stellar winds are taking 
place \citep{osk2005, naze2012}.  The lack of X-ray emission from single
LBVs  is expected  because their winds are quite slow, with  velocities
$\sim 100$\,km\,s$^{-1}$, and very dense, with $\dot{M}$ as high as 
$10^{-3}\,M_\odot$\,yr$^{-1}$ \citep{hill2001, Groh2009, Mary2012}. At 
such low wind velocities  wind shocks (even if  present) cannot produce
sufficiently hot plasma. Moreover, any X-rays produced in  the wind or
at the wind base would be fully absorbed \citep[e.g.][]{osk2005}.  

WR spectra are divided into three broad spectroscopic classes,  WN, WC,
and WO. WN stars display CNO-processed matter in a strong stellar  wind.
The cooler, late WN subtypes (WNL) usually contain some rest of
hydrogen  in their atmospheres, while the hotter, early subtypes (WNE)
are hydrogen free  \citep{ham1991}. The WNL evolutionary stage can
precede the LBV stage  \citep{Langer1994,Langer2012}. Typically, WNL
stars are significantly more  luminous than WNE stars
\citep{Hamann2006}. The WN phase may be followed by the  WC and WO
stages, when the products of helium burning appear in the stellar 
atmosphere \citep[see ][ for recent analyses]{Sander2012,Tramper2015}. 

Already from the analysis of early {\em Einstein} X-ray telescope 
observations it was found that binary WR stars are usually bright X-rays 
sources, while single WR stars are relatively faint \citep{Pollock1987}. 

In binary WR stars the bulk of X-rays is produced in  colliding winds. 
The  WN, WC, and WO type colliding wind binaries are  all bright X-ray
sources that usually show  variability  explained by orbital motion  
\citep[e.g.][]{Pollock2005,Sug2008,Lomax2015,Zhekov2012}. In some
objects,  besides X-rays,  non-thermal radio emission \citep{deb2013}
and dust are  produced \citep{Wil1990,Tut1999}.  

The X-ray spectra of  WR stars are well reproduced by multi-temperature
thermal plasma models with  temperatures  between 1\,MK up to 50\,MK 
\citep{Skina2002,Skin2010,Ignace2003,osk2012}.  The X-ray  production
mechanisms in these stars are not yet understood,  but it appears that 
X-ray production, like in O type stars, is an intrinsic ingredient of
the  stellar wind driving mechanism  (see K.\,Gayley, this volume). 

Yet there are significant differences in the X-ray properties of
different  types of WR stars.  The WNE stars have fast stellar winds,
similar effective  temperatures and luminosities \citep{Hamann2006}. All
well observed single  WN stars are X-ray sources with $L_{\rm X}
\approx  2\,...\,6 \times 10^{32}$\,erg\,s$^{-1}$ and an  emission
measure  weighted average temperatures of $\left<T\right>\approx
5$\,MK.  \citet{Chene2011,Chene-a2011} investigated the large-scale
spectroscopic  and photometric optical variability in apparently single
WR stars. Such  variability was found in four WN stars (WR1, WR6, WR110,
WR 134), and attributed  to the presence of CIRs. All these stars have
similar  X-ray properties. \citet{Chene2011} suggested as an explanation
for these quite  hard X-rays the shocks which are associated with the
CIRs.  

So far, a high-resolution X-ray spectrum is available for only one
single  WR star - EZ CMa (WR\,6) of spectral type WR4 \citep{osk2012,
Hue2015}.   Figure\,\ref{fig:zxmm} shows its RGS spectrum.  The resolved
X-ray lines are  broad and strongly blue-shifted as  expected when the
radiation forms in the outer wind and suffers strong continuum 
absorption \citep{Macfar1991,Ignace2001}. The analysis of line ratios
in  He-like ions confirms that the radiation is formed far above the
photosphere. Even  the hottest plasma is seen at more than ten stellar
radii. The fluorescent  Fe\,K$\alpha$ line is seen in the X-ray 
spectrum. This shows that dense and cool matter is illuminated by the
nearby X-ray sources. 

X-ray spectroscopy provides an excellent probe of metal abundances in stellar 
winds. Some of these abundances could not be measured  at 
other wavelengths. E.g., sodium lines were detected in the \cxo\ spectra of 
WR\,6. The spectral analysis revealed that the sodium abundance is enhanced in 
this star in accordance with its evolutionary state \citep{Hue2015}.

The WNL stars  are rather weak X-ray sources with X-ray luminosities not
exceeding   $L_{\rm X}\approx 10^{32}$\,erg\,s$^{-1}$ 
\citep{Wrigge1994,Pollock1995,Ignace2000,osk2005}. \citet{Skin2012}
reported a  2$\sigma$ detection of the WN8h star WR16 with $L_{\rm
X}\approx  10^{31}$\,erg\,s$^{-1}$. Some WNL stars were not detected in
X-rays despite  quite sensitive observations, e.g.\ \citet{Gosset2005}
found an upper limit to  the  X-ray luminosity of WR40 (WN8h) of $L_{\rm
X}< 4\times  10^{31}$\,erg\,s$^{-1}$. This is especially  puzzling,
because the wind of WR40  is more transparent for X-rays than the  wind
of WR6 (WN4), yet the latter is  $>100$ times more X-ray luminous than
the former. 

The WC stars are X-ray faint with  $L_{\rm X}/L_{\rm bol}< 10^{-9}$,
likely because of the high opacity of their dense and strongly metal
enriched winds \citep{osk2003}.  This conjecture is further supported by
the first  detection of faint X-ray emission from a single WC star.
\citet{Rauwa2015}  report the detection of WR144 with $L_{\rm X}\approx
10^{30}$\,erg\,s$^{-1}$. With its spectral type WC4, WR144 is one of
the hottest WC stars with the fastest wind among this class
\citep{Sander2012}. Thus its wind significantly less opaque for X-rays
than  in cooler WC types.  The intrinsic X-ray faintness of single WC
stars provides an  excellent  diagnostic  tool of binarity: X-ray bright
WC stars must be colliding wind systems
\citep[e.g.][]{Clark2008,osk2008,Hyodo2008, Mau2011,nebot2015}.

In WO winds the absorption is significantly reduced  compared to other
WR  stars, owing to lower wind column densities. Moreover, the very high
stellar  effective temperature of WO stars results in a higher degree of
wind ionization  compared to WN and WC stars, farther reducing the
opacity for X-rays  \citep{osk2012}. The wind speed in WO stars is very
high, up to  6000\,km\,s$^{-1}$ \citep{Sander2012,Tramper2015}. Hence,
{\em if} such stellar  wind is shocked, the temperature of the
heated plasma can reach up to  100\,MK. The resulting radiation would be
rather hard and would easier  escape from the  stellar wind. Thus, one
may expect that WO stars are detectable X-ray sources.   These
predictions were confirmed by the \xmm\ and \cxo\ detections  of hard
X-ray radiation from the WO2 star WR\,142 with $L_{\rm X} \approx 
10^{31}$\,erg\,s$^{-1}$ and $L_{\rm X}/L_{\rm bol}\sim 10^{-8}$
\citep{osk2009,Sokal2010}.  

Whatever is the mechanism of X-ray production in WR stars, the observed
X-ray  emission allows to probe their strong stellar winds. With the
notable  exception of WN8h type stars, the picture is emerging where the
X-ray  properties of WN, WC, and WO stars are chiefly determined by
their wind  opacity. The wind opacity of WC stars is higher than of WN
and WO stars. As a  result, we are able to see X-rays emerging from the
winds of WN and WO stars but not from WC type stars.    

\section{Summary}
\label{sec:sum}

Observations in the X-ray band of the electromagnetic spectrum provide
an important  diagnostic tool for stellar winds. In  early B-type and
late O-type dwarfs, the  bulk  of stellar wind may be  in a hot phase,
making the observations at X-ray wavelengths, the primary tool to  study
these most numerous massive stars. 

In OB giants and supergiants, the wind is mainly in the cool phase and
quite opaque for the X-rays.  Since the stellar winds are clumped, 
X-rays emitted from embedded sources can escape and reach the observer.
The  high-resolution X-ray spectra of OB supergiants allow detailed
studies of X-ray transfer in their winds. A brief parameter study shows
that the  fitting  of  X-ray line profiles by means of a simple model
does not provide a  reliable tool to measure stellar mass-loss rates. A
multiwavelength spectroscopic analysis, when the X-ray data are 
analyzed consistently with  optical and UV spectra, is required to 
derive realistic stellar and wind parameters.

X-ray variability emerges as a new powerful tool to understand stellar 
wind structures. Pronounced stochastic X-ray variability has not been
detected from stellar winds so far. However, all stars that were
observed with sufficiently  long exposures show X-ray variability on
time scales of days, similar to the  variability in H$\alpha$ and UV
lines. A likely mechanism for this  variability is by corotating
interaction regions.

Massive stars at the final stages of their evolution have diverse X-ray 
properties. The RSG stars appear to be X-ray dark.   Single LBV stars
and late sub-type WC stars are not observable as X-ray  sources, in
agreement with theoretical expectations.  Single  WN stars emit X-rays,
albeit the responsible mechanism is not yet  fully understood. The
hottest among all massive stars, the early sub-type WC and WO stars,
are  found to be X-ray sources, likely indicating that in these extreme
objects the attenuation of X-ray is sufficiently low.

\bigskip 
The author is indebted to both anonymous referees who provided 
important and detailed comments leading to a significant improvement  of
the manuscript. Support from DLR grant 50 OR 1302 is  acknowledged.

%\bibliography{xwind}
%\include{xlife.bbl}

\end{document}